# Noah's Flood and the Associated Tremendous Rainfall as a Possible Result of Collision of a Big Asteroid with the Sun.

by **Y. Y. Shopov[1], L. T. Tsankov[2], L. N. Georgiev[2], Y. Damyanov[2],**
[1]University Center for Space Research and Technologies,
[2]Faculty of Physics, University of Sofia, James Bourchier 5, Sofia 1164, Bulgaria.
E-mail: YYShopov@Phys.Uni-Sofia.BG
**A. Damyanova**
Institute of Astronomy of Bulgarian Academy of Sciences

**D.C. Ford**
School of Geology & Geography, McMaster University, Hamilton, Ontario, L8S 1K4, Canada

**C. J. Yonge,**
Alberta Karst Research Center, Kenmore, Alberta, Canada

## Abstract

A good correlation between the growth rate of the cave speleothems and the annual precipitation at the cave site allow quantitative reconstruction of the precipitation. Measuring the growth rate of a speleothem from Duhlata Cave, Bulgaria we found that around 7500 B.P. the speleothem growth rate (averaged for 120 years) exceeds 53 times its recent value suggesting that enormous precipitation flooded the Black Sea basin at that time. Its possible connection with the Bible (Noah's) Flood is discussed.

We propose a possible mechanism of the flooding of the Black Sea during the Flood involving production of a super- Tsunami by pushing of the Black Sea water towards the Crimea cost by Mediterranean waters.

We propose also an Astronomical Theory of the origin of the Bible Flood. We attribute higher water evaporation and rainfall to be caused by rapid increasing of the solar radiation resulting from a collision of a large asteroid or comet with the Sun.

## Method

Secondary cave calcite forms speleothems (stalactites, stalagmites, etc.) in caves. Speleothem luminescence visualises annual micro- banding under UV irradiation (Shopov et al. 1991, Shopov et al., 1994). We used it to derive proxy records of the annual precipitation at the cave site by measuring the distance between all adjacent annual maxima of the intensity of luminescence. The resultant growth rates correlate with the actual annual precipitation. We used it to derive quantitative proxy records of the annual precipitation at the cave site (Shopov et al., 1996, http://karst.planetresources.net/deluge.htm.). Dense dating of speleothems produce the same results for longer time span.

## Experimental Evidences for the (Bible Noah's) Flood

We measured variations of the growth rate of a speleothem from the Black Sea basin, Bulgaria representing past precipitation. at Bosnek karst region near Duhlata cave (DC), Bulgaria for the last 50000 years (Fig.1). This speleothem was dated with 8 TAMS $^{14}$C dates. This record shows a very prominent peak at 5500 years B.C., when the annual growth rate (averaged for 120 years) exceeds 53 times its recent value suggesting that enormous precipitation flooded the Black Sea basin at that time. Considering that cave site is located in the region of the oldest civilisations (Mediterranean basin) this event can be related to the Bible (Noah's) Flood or "Deluge" (Shopov et al., 1996, http://www.karst.edu.cn/igcp/igcp379/1997/part3-4-3.htm). The age of the recorded event is about the age of "The Creation of the World". The duration of the recorded event is of 120 years, because of the low resolution of the record. Presuming that the excess precipitation had fallen only within 1 year, this means a never seen rainfall (flood). Present day precipitation at the cave site is 650 mm/yr. Such event is described in the Bible, Greek mythology and the Sumerian epic Gilgamesh (compiled during 3-rd millennium B.C. on the base of more ancient legends). So probably it is the Noah's Flood. Calibrated AMS 14-C age of this event is the same as the beginning of the Bible chronology (Creation of the World), i. e. 5500 years B.C and also as the beginning of the Byzantine and Bulgarian calendars. At that time human civilisation had been concentrated around the Black and Mediterranean Sea, therefore the Flood hit also recent Bulgarian lands. Such immemorial precipitation probably would lead to some temporary rising of the Black Sea level. Such rising at 5500 B.C. with 150 meters (which had flooded 60000 square miles) was recently suggested by an international team of scientists, lead by Dr. William Rayn and Walter Pitman and confirmed by two expeditions of the National Geographic led by professor R. Ballard. The Black Sea level rising itself cannot be undoubtedly related to the Flood, but combined with the never seen (during the human civilisation) precipitation at that time definitely lead to the conclusion, that this phenomenon is namely the Bible Flood.

## Hypothesis for a Possible Mechanism of the Flood

We suggested a hypothesis for one possible mechanism of the Flood (Shopov et al., 1997), consisting of the following: - Ocean level raised from 10000 years B.P. to the time of the Flood as a result of the glaciers melting. Black Sea had been isolated from the Ocean and its level had been much lower. In one moment the narrow band of land between the Mediterranean and Black Sea had broken down like a dam wall. This had resulted in flowing of giant masses of seawaters into the Black Sea basin. When it reached the opposite cost a giant wave had been formed (which probably was incomparably bigger than the biggest tsunami known so far. This wave had destroyed everything on lands around the Black Sea even beyond the regions flooded by the sea level rise. The never seen precipitation at that time had contributed to the rising of the sea level and maybe caused the final rising, which had turned the Mediterranean Sea over the edge to flood the Black Sea region. Further studies of the Mediterranean Sea level during the Flood and data for precipitation from stalagmites taken from other caves in the region are necessary to prove this hypothesis.

## Astronomical Theory of the Noah's Flood

The mechanism of generation of such an enormously high rainfall is even more interesting. Any known Earth force is not able to produce such precipitation as the one measured by us and described in the historical sources. To generate such a rainfall is necessary to have enormous and rapid increasing in evaporation of the water, but there are no evidences of such rapid warming during the Flood. So the only possible reason for such evaporation is increasing of the solar luminosity with several %. Water absorbs strongly the infrared solar radiation, which cause melting of the glaciers and evaporation of the water. But evaporation cause cooling of the system (refrigerator effect). Solar luminosity usually remains rather steady. Such higher solar radiation can be produced only by explosion of a comet or an asteroid in the solar atmosphere. Such explosion (like Tungussian meteorite or the collision of parts of the Schumaker- Levy comet with Jupiter) can cause a major mixing of parts of the upper shells (layers) of the Sun and appearance of much warmer solar matter from the depth to the solar surface. Solar luminosity increases with the forth degree of the temperature of the solar surface, so it should increase significantly immediately after the collision (Shopov et al., 1997). Probably several years would be necessary for recovery of the Sun from similar catastrophic event.

Such rapid melting of the ice sheets and rising of the sea level should cause unusually rapid change of the rotation speed of the Earth and should produce major earthquakes. Probably one of them broke the narrow band of land separating Black Sea from the Mediterranean Sea and caused the flooding of the Black Sea basin.

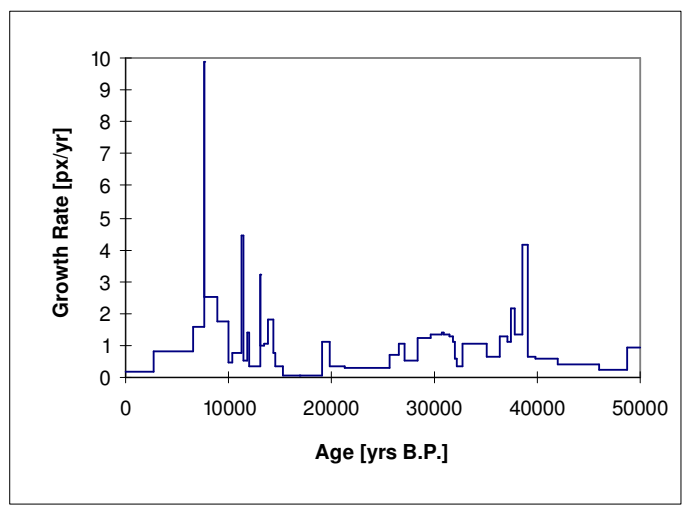

Fig.1. Reconstruction of the growth rates (proxy of precipitation) of a flowstone from DC, Bulgaria for the last 50000 years. (Shopov et al., 1996, Shopov et al., 1997)

## Conclusion

The Bible Flood (catastrophic rainfall) as recorded in a speleothem, had probably happened 5500 years B.C., causing Black Sea level rise. It is probably due to much higher water evaporation caused by rapid increasing of the solar radiation resulting from a collision of a large asteroid or comet with the Sun.

**Humanity even now is not prepared to face such catastrophic disaster so it is important to know its mechanism in order to help to predict it and to make proper actions to reduce the damage cause by it.**

## References

.Shopov Y.Y., et al, 1991: IGCP 299 Newsletter, 3: 52-58.
.Shopov Y.Y. et al., 1994: Geology, 22, 407-410.
.Shopov Y.Y., L.Tsankov, L.N.Georgiev, A.Damyanova, Y. Damyanov, E. Marinova, D.C. Ford, C.J.Yonge, W. MacDonald, H.P.R.Krouse (1996) Speleothem Luminescence proxy Records of Annual Rainfall in the Past. Evidences for "The Deluge" in Speleothems."- In book "Climatic Change- the Karst Record", Ed. by S.E. Lauritzen. Karst Water Research Institute. p. 155-156.

.Shopov Y.Y., L.Tsankov, L.N.Georgiev, A.Damyanova, Y. Damyanov, E. Marinova, D.C. Ford, C.J.Yonge, W. MacDonald, H.P.R.Krouse (1997) Evidences for "The Deluge" in Speleothems."- Proc. of 12th UIS Congress, La Chaux-de- Fonds, Switzerland, 10- 17 August 1997, v.1, pp.107- 109.